\documentclass[aps,twocolumn,groupedaddress]{revtex4}

\usepackage{graphicx}

\begin{document}


\title{Wireless communications with diffuse waves}


\author{S.E. Skipetrov}
\email[]{Sergey.Skipetrov@grenoble.cnrs.fr}
\affiliation{CNRS/Laboratoire de Physique et Mod\'elisation des Milieux Condens\'es,\\
38042 Grenoble, France}


\date{\today}

\begin{abstract}
Diffuse, multiple-scattered waves can be very efficient for information transfer
through disordered media, provided that antenna arrays are used for both
transmission and reception of signals.
Information capacity $C$ of a communication channel between
two identical linear arrays of $n$ equally-spaced antennas,
placed in a disordered medium with diffuse scattering,
grows linearly with $n$ and can attain considerable
values, if antenna spacing $a \agt \lambda/2$, where $\lambda$ is the wavelength.
Decrease of $a$ below $\lambda/2$
makes the signals received by different antennas partially
correlated, thus introducing redundancy and reducing capacity of the
communication system. When the size of antenna arrays is well below
$\lambda/2$, the scaling of $C$ with $n$ becomes logarithmic and capacity
is low.
\end{abstract}


\maketitle

Wireless communications in a disordered environment have
recently attracted considerable attention \cite{mous00,andrews01,simon01}.
At first glance, scattering from randomly distributed heterogeneities disturbs
the signal carried by the scattered wave (either acoustic or electromagnetic) and
is expected to
reduce the efficiency of communication.
Indeed, the rate
of error-free information transfer per Hertz of frequency bandwidth
for a scalar, narrow-band communication channel between a transmitter and
a receiver is bounded by the \textit{channel information capacity}
$= \log_2 [ 1 + \left| G \right|^2 / N ]$
measured in bits per second per Hertz (bps/Hz) \cite{shannon48,cover91}.
Here $G = G(\nu)$ is the Fourier transform of the impulse response function $G(t)$,
describing the signal at the receiver due to an infinitely short pulse emitted by the
transmitter, and $N$ is the power of the Gaussian white noise at the receiver.
Obviously, the decrease of the signal power $\left| G \right|^2$ due
to scattering results in a reduction of the channel capacity.
However,
in addition to the overall decrease of $\left| G \right|^2$, scattering introduces
random fluctuations of $G$ in space.
It was recently realized \cite{foschini98} that one can make use of these fluctuations
to overcome the reduction of $C$ due to the drop of the signal power,
provided that the communication system contains a large number of antennas
\cite{mous00,simon01}. This issue is very promising for applications in wireless communications
(mobile telephony in cities, indoor wireless local-area networks in
buildings with complex structure,
underwater communications with acoustic waves, etc.).

Assuming that the information about
the scattering environment is available at the receivers (but not at the transmitters)
and that uncorrelated noises at different receiving antennas have the same power $N$,
the average capacity of a communication channel between arrays of
$n$ transmitting and $n$ receiving antennas \footnote{For simplicity, we consider
the transmitting and receiving arrays to consist of the same number $n$ of antennas.}
can be defined as \cite{mous00}
\begin{eqnarray}
C =
\max\limits_{\mathbf{Q}} \left< \log_2 \det \left[ \mathbf{I}_n + \mathbf{G} \,
\mathbf{Q} \, \mathbf{G}^+ / N \right]
\right>,
\label{c}
\end{eqnarray}
where
$\mathbf{I}_n$ is the $n \times n$ unit matrix, $\mathbf{G}$
is a Green matrix ($G_{i \alpha}$
gives the signal received by the receiver $\alpha$ due to the transmitter $i$),
and $\mathbf{Q}$ is a non-negative definite covariance matrix
describing correlations between the transmitted signals (with the
constraint on the maximum transmitted power $\mathrm{Tr}\, \mathbf{Q} \leq n$).
Angular brackets in Eq.\ (\ref{c}) denote averaging over realizations of disorder.
A rigorous analysis of the information capacity $C$ in a disordered medium
cannot be based uniquely on the arguments of the information theory
\cite{shannon48,cover91} and requires
the physical understanding of scattering undergone by the waves carrying the
information from transmitters to receivers. In the present paper we
analyze $C$ in the framework of the model of diffuse multiple scattering, resulting
in a complicated, seemingly random spatial distribution
of scattered wave fields (so-called ``speckles'').
For the extreme cases of (a) a single speckle spot covering the whole antenna array
($\mathbf{G}$ has perfectly correlated entries) and (b) different antennas
situated inside different speckle spots (entries of $\mathbf{G}$ are uncorrelated),
we derive complete analytical expressions for $C$.
We show that $C$ increases during the continuous
transition from the case (a) to the case
(b) upon increasing antenna spacing (or, equivalently, upon decreasing
correlations between the entries of $\mathbf{G}$) and
changes its asymptotic scaling with $n$ from $C \propto \ln n$
to $C \propto n$.

To be specific, we consider two identical linear arrays of equally-spaced
transmitting/receiving antennas placed in a disordered medium.
The distance $L$ between the arrays is assumed to be much greater than the
mean free path $\ell$ for waves in the medium, while the array size
$d = (n-1) a \ll \ell$ and $d \ll (\lambda L)^{1/2}$ (Fresnel limit),
where $a$ is the spacing between adjacent
antennas and $\lambda$ is the wavelength.
In the majority of practically important cases $\lambda \ll \ell$,
and hence the propagation of waves from transmitters to
receivers is diffusive \cite{ishim78}.
Entries $G_{i \alpha}$ of the matrix
$\mathbf{G}$ can be then treated as complex Gaussian random variables with zero mean,
equal variances,
and possibly nontrivial correlations \cite{shapiro86}
\begin{eqnarray}
\left< G_{i \alpha} G_{j \beta} \right> =
\sigma^2 \frac{\sin(k \Delta r_{i j})}{k \Delta r_{i j}}
\frac{\sin(k \Delta r_{\alpha \beta})}{k \Delta r_{\alpha \beta}},
\label{corr}
\end{eqnarray}
where
$k = 2 \pi / \lambda$,
$\Delta r_{i j} = \left| i - j \right| a$,
and $\Delta r_{\alpha \beta} = \left| \alpha - \beta \right| a$.

\begin{figure}
\includegraphics[width=8cm]{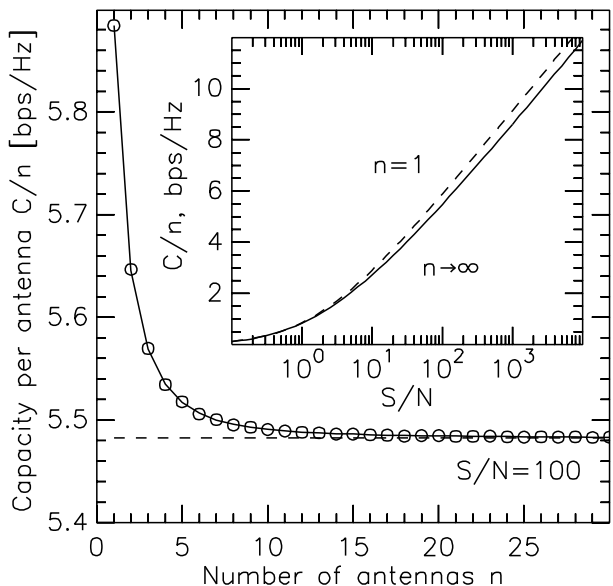}
\caption{\label{fig1} Average information capacity per antenna of
a communication channel between two arrays of $n$ antennas in a disordered
medium assuming statistical independence of the entries of the Green
matrix $\mathbf{G}$:
results obtained using the random matrix theory (solid line)
and Monte Carlo simulation (symbols) are shown for the signal to noise ratio
$S/N = 100$. Horizontal dashed line shows the asymptotic
value of $C/n$ for $n \rightarrow \infty$. Inset: Capacity per antenna as
a function of $S/N$ for single transmitting and single receiving antennas
(dashed line) and an infinitely large number of antennas (solid line).}
\end{figure}

The simplest cases to consider are (a) $k d \rightarrow 0$ and
(b) $k a = m \pi$ ($m = 1, 2, \ldots$).
In the former case, all $G_{i \alpha}$ are
perfectly correlated and one finds
$C = \exp[1/(n^2 S/N)] E_1[1/(n^2 S/N)] / \ln 2$,
where $E_1$ is the exponential integral function,
$S = n \sigma^2$
is the average signal power received at each receiver assuming
independent signals from transmitters, and
$C \propto \ln n$ for $n^2 S/N \gg 1$.
In the case (b),
$\mathbf{Q} = \mathbf{I}_n$ in Eq.\ (\ref{c})
\cite{telatar99} and
the machinery of the random matrix theory \cite{mehta91} can be employed
for further analysis of the problem. In particular, it is useful to rewrite
Eq.\ (\ref{c}) as
\begin{eqnarray}
C = \left< \sum\limits_{i = 1}^n \log_2 \left[ 1 + (S/N) \mu_i \right] \right>,
\label{c1}
\end{eqnarray}
where $\mu_i$ are the squares of the singular values of the matrix
$\mathbf{G}^+/ S^{1/2}$
with the joint probability density function \cite{mehta91}
\begin{eqnarray}
p_n(\mu_1, \ldots, \mu_n) = A_n
\exp\left( -n \sum\limits_{i=1}^n \mu_i \right)
\prod\limits_{i < j} (\mu_i - \mu_j)^2,
\label{rmt}
\end{eqnarray}
where $A_n$ is a normalization constant. The moment generating function $F(x)$ of
a random variable $c$, defined by Eq.\ (\ref{c1}) without
averaging, is obtained by averaging
$\exp(x c)$ with help of Eq.\ (\ref{rmt}) \cite{sengupta00},
and $C = \left< c \right>$ is then calculated as
\begin{eqnarray}
C = \frac{\mathrm{d}}{\mathrm{d} x} F(x) \Big|_{x = 0} =
\mathrm{Tr} \left( \mathbf{P}^{-1} \mathbf{R} \right),
\label{c2}
\end{eqnarray}
where $\mathbf{P}$ and $\mathbf{R}$ are $n \times n$ matrices:
\begin{eqnarray}
P_{i j} &=& (i + j - 2)!\, n^{1-i-j},
\label{p} \\
R_{i j} &=& \int_0^{\infty} \mathrm{d} \mu \,
\ln\left[ 1 + (S/N) \mu \right] \mu^{i+j-2} \exp(-n \mu).
\label{q}
\end{eqnarray}
Equations (\ref{c2})--(\ref{q}) provide an efficient way for capacity calculation
at arbitrary $n$ and $S/N$ (see the solid line in Fig.\ \ref{fig1}) and agree
perfectly with the direct Monte Carlo simulation of capacity using
Eq.\ (\ref{c}) (symbols in Fig.\ \ref{fig1}).
Note that the growth of $C$ with $n$ is linear at large $n$ and hence is
much faster than in the case (a).

An alternative way of calculating capacity consists in performing averaging
directly in Eq.\ (\ref{c1}) using Eq.\ (\ref{rmt}). This yields
\begin{eqnarray}
C/n = \int_0^{\infty} \mathrm{d} \mu \log_2
\left[ 1 + (S/N) \mu \right] f_n(\mu),
\label{c3}
\end{eqnarray}
where
$f_n(\mu_1) = \int_0^{\infty} \mathrm{d} \mu_2 \ldots \mathrm{d} \mu_n \,
p_n(\mu_1, \ldots, \mu_n)$ can be evaluated by a direct integration, at least at
moderate $n$:
$f_1(\mu) = \exp(-\mu)$,
$f_2(\mu) = 2 \exp(-2 \mu) [1 - 2 \mu + 2 \mu^2]$,
$f_3(\mu) = 3 \exp(-3 \mu) [1 - 6 \mu + 18 \mu^2 - 18 \mu^3 + (27/4) \mu^4]$,
etc.
The values of capacity obtained then from Eq.\ (\ref{c3}) coincide exactly
with that following from Eq.\ (\ref{c2}).
An asymptotic expression for $f_n(\mu)$ at $n \gg 1$ can be
found in the framework of the random matrix theory \cite{mehta91,sengupta99}:
$f_{\infty}(\mu) = (2 \pi)^{-1} (4/\mu - 1)^{1/2}$
for $0 < \mu < 4$ and $f_{\infty}(\mu) = 0$ otherwise. Eq.\ (\ref{c3}) then yields
\begin{eqnarray}
C/n &=&
(S/N) \,_3F_2 \left( 1, 1, 3/2;\, 2, 3;\, -4\, S/N \right) / \ln 2,
\label{cinf}
\end{eqnarray}
where $_3F_2$ is the generalized hypergeometric function.
This result is shown in
the main plot of Fig.\ \ref{fig1} by a dashed horizontal line.
We find $C/n \simeq \log_2(S/N) - \log_2 e$ for $S/N \gg 1$
(see also the inset of Fig.\ \ref{fig1}).
It is worthwhile to note that $C/n$ decreases monotonically with $n$,
while the difference between the values of $C/n$ at $n=1$ and $n > 1$
never exceeds 7\%.

\begin{figure}
\includegraphics[width=8cm]{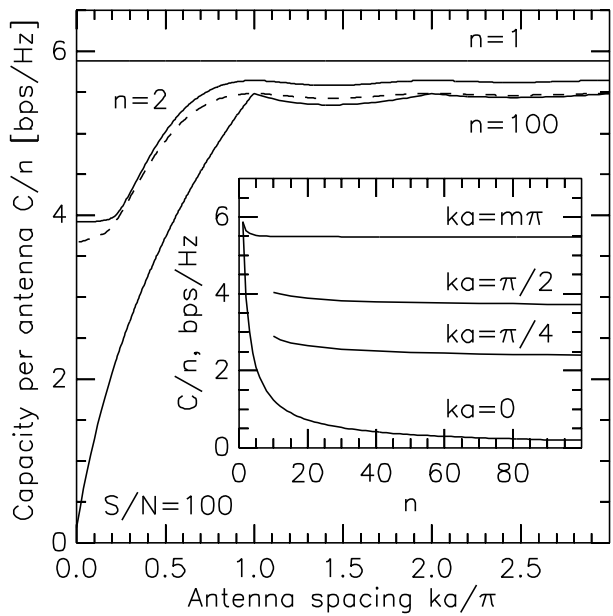}
\caption{\label{fig2} Average information capacity per antenna of
a communication channel between two identical linear arrays of $n$ antennas
in a disordered medium assuming antenna spacing $a$ and $S/N = 100$.
Solid lines correspond to $n=1$ (exact result), $n=2$ (Monte Carlo simulation)
and $n=100$ (asymptotic large-$n$ result).
Dashed line shows the result obtained using the asymptotic
large-$n$ theory at $n=2$.
Inset: Capacity per antenna as
a function of $n$ for different antenna spacings $a$. Results for
$ka = \pi/2$ and $\pi/4$ are asymptotic large-$n$ results.}
\end{figure}

We now allow $k a$ to take arbitrary values, thus introducing
correlations between the entries of the Green matrix $\mathbf{G}$.
Eq.\ (\ref{c}) can again be reduced to Eq.\ (\ref{c1})
with $\mu_i$ denoting the squares of the singular values of the matrix
$\mathbf{Q}^{1/2} \mathbf{G}^+/ S^{1/2}$,
where $\mathbf{Q}$ is chosen to
maximize the result.
The joint probability density function of $\mu_i$,
analogous to Eq.\ (\ref{rmt}), is not known in this case,
but one can still calculate $C$ in
the large-$n$ limit \cite{mous00,sengupta00}.
The idea is to represent the moment generating function $F(x)$
of a random variable $c$, defined by Eq.\ (\ref{c1}) without averaging,
as a multiple Gaussian integral
(the so-called ``replica trick'') and then to perform the integrations using
saddle point methods in the limit $n \gg 1$. The maximization
of $C = (\mathrm{d} / \mathrm{d} x) F(x) |_{x = 0}$
over
$\mathbf{Q}$ is then
accomplished by requiring $\delta C \leq 0$ for all allowed small variations
$\delta \mathbf{Q}$ of the optimal matrix $\mathbf{Q}$.
This yields a system of nonlinear equations for the eigenvalues of
$\mathbf{Q}$ and some auxiliary variables that can be solved numerically.
We refer the reader to Refs.\ \onlinecite{sengupta00} and \onlinecite{sengupta99}
for an exhaustive account of the theoretical approach and the algorithm of
the numerical calculation.
The results that we obtained for identical linear arrays of equally-spaced
antennas are presented in Fig.\ \ref{fig2}.
As follows from the figure, at $ka > \pi$ correlations between
the entries of $\mathbf{G}$ are too weak to affect $C$ significantly and
the latter remains very close to its maximum value, given by Eqs.\ (\ref{c2})
or (\ref{c3}).
In contrast, correlations become important
when $ka$ decreases below $\pi$, leading to a significant drop of $C$.
The dashed line in Fig.\ \ref{fig2} results from the asymptotic large-$n$ theory
\cite{mous00,sengupta00,sengupta99} with $n = 2$ and is shown for
illustration purposes only. Its closeness to the Monte Carlo result
(solid line for $n=2$) testifies a qualitative validity of the large-$n$ theory
even at small $n$.
As $k a$ increases, the scaling of $C$ with $n$ changes from $C \propto \ln n$ at
$ka=0$ to $C \propto n$ at $ka \agt \pi$
(see the inset of Fig.\ \ref{fig2}). We note that even
at $0 < ka < \pi$,
there is still a significant gain in capacity as compared to the case of
$ka = 0$: e.g., at $n = 100$, $C/n$ for $ka = \pi/2$ ($\pi/4$)
is almost 20 (13) times larger than for $ka = 0$.

In conclusion, we have studied the information capacity $C$ of a wireless
communication channel in a disordered medium, assuming multiple diffuse scattering
of waves that carry information, and taking into account fluctuations of
wave fields in space (speckles).
Although multiple scattering reduces
the received signal, it allows for a dramatic increase of capacity in the case of
communication between two antenna arrays, provided that antenna
spacing $a \agt \lambda/2$, where $\lambda$ is the wavelength. Namely,
for identical linear arrays of $n$ equally spaced receiving/transmitting antennas,
scaling of $C$ with $n$ changes from $C \propto \ln n$ for $kd \ll \pi$
to $C \propto n$ for $ka \agt \pi$, where $d = a (n-1)$ is the array size and
$k = 2 \pi / \lambda$. Even at $0 < ka < \pi$ an important
gain in capacity is possible as compared to $ka = 0$.

The author is indebted to Prof. R. Maynard for helpful discussions
and critical reading of the manuscript.

\end{document}